\documentclass[twocolumn,showpacs,amsmath,amssymb]{revtex4}
\usepackage{graphicx}
\usepackage{dcolumn}
\usepackage{bm}
\newcommand{\be}{\begin{eqnarray}}
\newcommand{\en}{\end{eqnarray}}
\newcommand{\ben}{\begin{eqnarray*}}
\newcommand{\enn}{\end{eqnarray*}}
\newcommand{\pa}{\partial}
\newcommand{\na}{\nabla}
\newcommand{\f}{\frac}

\newcommand{\bi}{\begin{itemize}}
\newcommand{\ei}{\end{itemize}}
\newcommand{\im}{\item}

\begin{document}
\title{On knotted streamtubes in incompressible hydrodynamical flow and a restricted conserved quantity}
\author{Sagar Chakraborty}
\email{sagar@bose.res.in}
\affiliation{S.N. Bose National Centre for Basic Sciences\\ Saltlake, Kolkata 700098, India}
\date{February 26, 2006}
\begin{abstract}
For certain families of fluid flow, a new conserved quantity -- stream-helicity -- has been established.
Using examples of linked and knotted streamtubes, it has been shown that stream-helicity does, in certain cases, entertain itself with a very precise topological meaning {\it viz.}, measure of the degree of knottedness or linkage of streamtubes.
As a consequence, stream-helicity emerges as a robust topological invariant.
\end{abstract}
\maketitle
\section{INTRODUCTION}
Lord Kelvin (who alongwith Helmholtz pioneered in the subject of vortex motion) had recognised in the late 19th century that in an inviscid and barotropic fluid being acted upon by irrotational body forces, any linkage or any knottedness in the vorticity field at any earlier time should remain conserved at all later times.
After almost hundred years, Moreau\cite{1} and later Moffatt\cite{2} established an invariant known as helicity which is of topological character and encompasses Kelvin's insight.
Stark analogy between vorticity ($\vec{\omega}$) in ordinary fluid dynamics and magnetic field ($\vec{B}$) in magneto-hydrodynamics (MHD) prompted Moffatt to give similar topological interpretations to magnetic helicity and cross-helicity (which, by the way, measures the degree of `mutual' knottedness of two the fields: $\vec{\omega}$ and $\vec{B}$).
%
%
Hence, researchers were able to effectively connect the two very rich fields {\it viz.}, topology and fluid dynamics and excited a lot of interest in this direction. 
But what lord Kelvin had missed was the possible existence of knotted streamtubes in the steady Euler flows, a fact very logically speculated by Moffatt\cite{3}.
Not much has been done on that.
Here, in this paper, inspired by the works of Moffatt, we shall introduce in section (II) a quantity, which we shall call `stream-helicity' ($S$) in inviscid and incompressible fluid being forced by irrotational body forces.
It will be shown that stream-helicity is a conserved quantity under certain restrictions which are not, of course, very rare in practice.
In section (III), we shall note how this conserved quantity can have a very sound topological meaning for at least some kinds of flows and hence, how stream-helicity can be raised to the status of a topological invariant for linked and knotted streamtubes.
\section{STREAM-HELICITY}
Let us start with the Euler equation (equation (\ref{1})) for three-dimensional, inviscid, incompressible fluid being acted upon by irrotational body forces.
$P$ used in the equation includes the effect of such forces also.
Since the fluid is incompressible, i.e., the density is constant, we are taking the density to be unity for convenience.
\be
\f{\pa}{\pa t}{\vec{u}}+({\vec{u}}.{\vec{\na}}){\vec{u}}=-{\vec{\na}}P
\label{1}
\en
Incompressibility yields for the velocity field $\vec{u}$:
\be
{\vec{\na}}.{\vec{u}}=0
\label{2}
\en
which helps in defining the vector potential $\vec{\xi}$ for the velocity field as follows:
\be
{\vec{u}}={\vec{\na}}\times{\vec{\xi}}
\label{3}
\en
Obviously, $\vec{\xi}$ is not unique, for, a term $\vec{\na}\lambda$, $\lambda$ being a scalar field can always be added to it keeping $\vec{u}$ unchanged.
We shall come back to this issue in the right place.
For now, let us put relation (\ref{3}) in the equation (\ref{1}) to get:
\be
\f{\pa}{\pa t}({\vec{\na}}\times{\vec{\xi}})+({\vec{u}}.{\vec{\na}})({\vec{\na}}\times{\vec{\xi}})=-{\vec{\na}}P
\label{4}
\en
But we have:
\be
[{\vec{\na}}\times({\vec{u}}.{\vec{\na}}){\vec{\xi}}]_i&=&\epsilon_{ijk}\pa_j(u_l\pa_l\xi_k)\nonumber\\
&=&\epsilon_{ijk}(\pa_ju_l)(\pa_l\xi_k)+\epsilon_{ijk}u_l\pa_j\pa_l\xi_k\nonumber\\
&=&\epsilon_{ijk}(\pa_ju_l)(\pa_l\xi_k)+[({\vec{u}}.{\vec{\na}})({\vec{\na}}\times{\vec{\xi}})]_i
\label{5}
\en
Using relation (\ref{5}) in the equation (\ref{4}) we get:
\be
&{}&\left[\f{\pa}{\pa t}({\vec{\na}}\times{\vec{\xi}})+{\vec{\na}}\times\{({\vec{u}}.{\vec{\na}}){\vec{\xi}}\}\right]_i=\epsilon_{ijk}(\pa_ju_l)(\pa_l\xi_k)-\pa_iP\nonumber\\
\Rightarrow\phantom{x}&{}&\f{\pa}{\pa t}{\vec{\xi}}+({\vec{u}}.{\vec{\na}}){\vec{\xi}}=\textrm{curl}^{-1}{\vec{\eta}}
\label{6}
\en
where $\vec{\eta}$ is defined as:
\be
{\eta}_i\equiv\epsilon_{ijk}(\pa_ju_l)(\pa_l\xi_k)-\pa_iP
\label{7}
\en
Now, let us define `stream-helicity' ($S$) as:
\be
S\equiv\int_V{\vec{\xi}}.{\vec{u}}d^3x
\label{8}
\en
where $V$ is a volume occupied by the fluid.
At this point let us ponder over the aforementioned non-uniqueness of the vector potential\cite{Berger}.
For smooth discussion's sake, we assume for the time being that the volume is simply connected.
Suppose $\xi_i\rightarrow\xi_i+\pa_i\lambda$, then from the definition (\ref{8}) of stream-helicity we can find the change $\delta{S}$ in $S$ to be:
\be
\delta S=\int_V\vec{\na}\lambda.\vec{u}d^3x=\oint_{\pa{V}}\lambda\vec{u}.\hat{n}d^2x
\label{ds}
\en
where $\hat{n}$ is the unit vector perpendicular to the infinitesimal surface element $d^2x$ and we have used the relation (\ref{2}) and Gauss divergence theorem.
The relation (\ref{ds}) amounts to saying that the stream-helicity will be gauge invariant in case the surface $\pa V$ bounding $V$, is the surface made up of streamlines {\it i.e.}, $\vec{u}.\hat{n}=0$ on $\pa V$.
This condition for gauge invariance is rather strong because if $\vec{u}.\hat{n}\ne 0$ on $\pa V$ then one cannot seek refuge in Coulomb gauge for it is too loosely defined inside $V$ with no information about the outside field whatsoever.
More starkly, it means that different solenoidal vector potentials inside $V$ can correspond to Coulomb potentials of fields which have different structures outside $V$.
Now if we relax the condition that $V$ is simply connected, then the line integrals of $\vec{\xi}$ about the `holes' in the possibly multiply connected region have to be specified in order to have gauge-invariant stream-helicity within $\pa V$ on which $\vec{u}.\hat{n}=0$.
The upshot is that only in the gauge $\vec{u}.\hat{n}=0$ (which in the simple language means that the boundary is impenetrable), $S$ can be conserved.
\\
We now wish to demonstrate that under certain restrictions this quantity is in fact conserved.
So, we take total derivative of $S$ w.r.t. time to get:
\be
\f{dS}{dt}&=&\int\f{D}{Dt}({\vec{\xi}}.{\vec{u}})d^3x\nonumber\\
\Rightarrow\f{dS}{dt}&=&\int{\vec{\xi}}.(-{\vec{\na}}P)d^3x+\int{\vec{u}}.(\textrm{curl}^{-1}{\vec{\eta}})d^3x
\label{9}
\en
where, $D/Dt$ is the material derivative w.r.t. time and it basically is a shorthand for $\pa/\pa t+\vec{u}.\vec{\na}$.
Again, simple vector algebra suggests:
\be
({\vec{\na}}\times{\vec{\xi}}).(\textrm{curl}^{-1}{\vec{\eta}})={\vec{\na}}.({\vec{\xi}}\times\textrm{curl}^{-1}{\vec{\eta}})+{\vec{\eta}}.{\vec{\xi}}
\label{10}
\en
With relation (\ref{3}) in mind, inserting relation (\ref{10}) in the equation (\ref{9}), we have the following:
\be
\f{dS}{dt}&=&-2\int{\vec{\xi}}.{\vec{\na}}Pd^3x+\int{\vec{\na}}.({\vec{\xi}}\times\textrm{curl}^{-1}{\vec{\eta}})d^3x\nonumber\\
&&+\int\xi_i\pa_l(\epsilon_{ijk}\xi_k\pa_ju_l)d^3x
\label{11}
\en
where, equation (\ref{2}) has been used.
The first two terms of the equation (\ref{11}) can be changed to integration over the surface which bounds the volume $V$ in question (the surface will obviously extend to infinity if the fluid is unbounded) using Gauss divergence theorem and so if $\vec{\xi}$ decays fast enough to go to zero on the bounding surface then these two term vanish; there may be other reasons for the terms to vanish as will be seen in the next section.
Now, let us consider the third term.
If this term vanishes then only one may set
\be
\f{dS}{dt}=0
\label{12}
\en
and say that stream-helicity is a conserved quantity.
Though it seems to be very restrictive, but one can see that in the following commonly occurring cases the integrand of this term trivially vanishes.
\begin{enumerate}
\im The vector potential is one dimensional.
\im The vector potential has no dependence on the direction along the velocity field. (Other conditions given below are basically this condition's corollary.)
\begin{enumerate}
\im $\vec{\xi}$ is two dimensional but has dependence only on the third direction.
\im $\vec{\xi}$ is two dimensional with spatial variations only on the plane containing it.
\im $\vec{\xi}$ is three dimensional but depends only on any one of the three independent directions.
\end{enumerate}
\end{enumerate}
One can see that such flows are very commonly found in any elementary text-books on fluid mechanics.
For example, in accordance with case (1), for the one dimensional vector potential: $\vec{\xi}=\hat{k}\Omega(x^2+y^2)/4$, the corresponding flow is $\vec{u}=-\hat{i}\Omega y/2+\hat{j}\Omega x/2$ which basically is the velocity field for three dimensional fluid counter-clock-wisely rotating about z-axis.
As another instance, this time to go with the case (2) (2(b), to be precise), is that of a uniform flow along x-direction: $\vec{u}=U\hat{i}$ which is generated by the vector potential $\vec{\xi}=-\hat{j}Uz/2+\hat{k}Uy/2$ that evidently is two dimensional with spatial variations only on the y-z plane containing it.
A rather non-trival case (as an example of the case 2(a)) is for the flow: $\vec{u}=\hat{i}U\sin{az}+\hat{j}U\cos{az}$ ($a$ is a constant) for which the vector potential is $\vec{\xi}=\hat{i}(U/a)\sin{az}+\hat{j}(U/a)\cos{az}$; readers must have noticed that this is just a variant of the more general flow {\it {viz.}}, ABC flow (see {\it e.g.}, \cite{Arnold}).
\\
So, for the families of fluid flow for which the vector potential falls into the above set and if eventually the equation (\ref{12}) holds, stream-helicity is a conserved quantity.
Also, for the fluid flows for which doesn't fall in the above set but the integration goes to zero some reason or the other (which has not been investigated), $S$ will remain conserved.
\section{TOPOLOGICAL MEANING OF STREAM-HELICITY}
Now, we ask the question if it is possible to give stream-helicity a topological meaning and more importantly, can that topological meaning turn out to be a topological invariant.
We shall see that the answer is in affirmative.
To get both of the expectations met, one (other uninvestigated possibilities may also be there) of the ways seems to be the following:\\
Consider two circular thin streamtubes which are singly-linked and for the two tubes the `strengths' are $V_{C_1}$ and $V_{C_2}$ respectively, where $C_1$ and $C_2$ denote axis-circles of the corresponding tubes.
By `strengths' we mean that $V_{C_1}=\vec{u}.d\vec{a}_1$ and $V_{C_2}=\vec{u}.d\vec{a}_2$.
(See FIG-1).
\begin{figure}
\centering
\includegraphics[width=7.0cm]{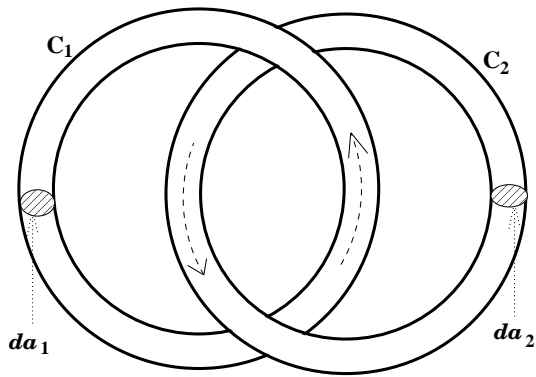}
\caption{Linked closed streamtubes. The tube and hence the streamlines inside are not twisted {\it i.e.}, the fluid inside the tube doesn't swirl. The directions of arrows are showing the direction of the streamlines filling the tubes.}
\end{figure}
Again, we assume that the velocity field, we shall be dealing with, is generated by a vector potential $\vec{\xi}$ which is a Beltrami field i.e.,
\be
\vec{u}=\vec{\na}\times\vec{\xi}=\alpha\vec{\xi}
\label{13}
\en
where, $\alpha$ is a numerical constant.
Moreover, suppose that of the conditions gathered in the previous section for $\vec{\xi}$, at least one is applicable, say the second one that:
\be
({\vec{u}}.{\vec{\na}}){\vec{\xi}}={\vec{0}}
\label{14}
\en
whether this is possible or not may be a valid question.
One may derive `some' relief from the fact that if $\vec{\xi}$ is analogous to ABC flow (Gromeka(1881); Beltrami(1889)), then it does satisfy such condition though unfortunately it may not sustain a linked structure of streamtubes.
Then the streamtubes will be made up of streamlines which are coincident with the `flux-lines' of $\vec{\xi}$ field.\\
If we define the volume over which the integration is defined for the stream-helicity to be the volume occupied by the linked structure only, then
\be
S&\equiv&\int{\vec{\xi}}.{\vec{u}}d^3x=\int\int\int_{\textrm{1st Streamtube}}{\vec{\xi}}.{\vec{u}}d^3x+\nonumber\\
&&\int\int\int_{\textrm{2nd Streamtube}}{\vec{\xi}}.{\vec{u}}d^3x\nonumber\\
\Rightarrow S&=&V_{C_1}\int_{C_1}{\vec{\xi}}.d{\vec{l}}_1+V_{C_2}\int_{C_2}{\vec{\xi}}.d{\vec{l}_2}\nonumber\\
\Rightarrow S&=&V_{C_1}\int\int_{DC_1}{\vec{u}}.d{\vec{\sigma}}+V_{C_2}\int\int_{DC_2}{\vec{u}}.d{\vec{\sigma}}\nonumber\\
\Rightarrow S&=&V_{C_1}V_{C_2}+V_{C_1}V_{C_2}\nonumber\\
\Rightarrow S&=&2V_{C_1}V_{C_2}
\label{15}
\en
where in the preceding steps we have used $\vec{u}d^3x\rightarrow V_{C_1}d\vec{l}_1$, $V_{C_2}d\vec{l}_2$ on $C_1$ and $C_2$ respectively; $DC_1$ and $DC_2$ denote the area spanned by $C_1$ and $C_2$ respectively.
Obviously, if the linking number is $n$ and not one as in this case, one would easily generalise the result to:
\be
S&=&2nV_{C_1}V_{C_2}
\label{16}
\en
which, being dependent on the mutual linking of streamtubes, is a topological quantity.
One may write from equation (\ref{11}) using Gauss divergence theorem in the following form:
\be
\f{dS}{dt}&=&-2\int({\vec{\xi}.\hat{n}})Pd^2x+\int(\textrm{curl}^{-1}\vec{\eta}).(\hat{n}\times{\vec{\xi}})d^2x\nonumber\\
&&+\int\xi_i\pa_l(\epsilon_{ijk}\xi_k\pa_ju_l)d^3x
\label{17}
\en
where $\hat{n}$ is the unit vector perpendicular to the surface at each point on the surface of the linked structure.
The first term and the third terms of the equation ({\ref{17}}) are zero in this case by construction of the linked structure; so is the second term but it needs a bit of manipulation as explained below.\\
First of all, we use equation (\ref{6}), to rewrite the integrand of the second term of the relation (\ref{17}) as
\be
\left[\f{\pa{\vec{\xi}}}{\pa t}+({\vec{u}}.{\vec{\na}}){\vec{\xi}}\right].(\hat{n}\times{\vec{\xi}})=\epsilon_{ijk}n_j\xi_k\left[\f{\pa{{\xi}_i}}{\pa t}+({{u}_l}{{\pa}_l}){{\xi}_i}\right]
\label{18}
\en
Now, if we consider the Frenet-Serret coordinate system: $(\vec{T},\vec{N},\vec{B})$, then in the case we are considering $\vec{\xi}/|\vec{\xi}|=\vec{T}$ and $\hat{n}=\vec{N}$; obviously on the surface of the specific tube we are considering, at each point, the triad so that there $\xi_2=\xi_3=0$ and $n_1=n_3=0$ and hence due to the antisymmetry of $\epsilon_{ijk}$ we have
:%
\be
(\textrm{curl}^{-1}\eta).(\hat{n}\times{\vec{\xi}})=0
\label{19}
\en
So, obviously we land up on the following relation:
\be
\f{dS}{dt}=0
\label{20}
\en
Therefore, for incompressible, ideal and conservatively forced fluid flow, in certain configurations, we can have a topological invariant -- stream-helicity -- for linked structures of streamtubes.\\
So far so good.
So, stream-helicity does seem to make physical sense for linked two (or more) streamtubes.
But what if a single streamtube is knotted?
A single knotted streamtube must have an unavoidable twist of velocity field (which we hope, in this case also, may be derived from a Beltrami velocity vector potential and is of similar kind as has been dealt with earlier in this paper) within the tube.
How to deal with such a scenario has been discussed for knotted vortex filaments by Moffatt\cite{4}.
We know when an arbitrary tame knot is viewed in a standard plane of projection with finite number of crossings, each of which is either positive or negative, it can be changed to a unknot (and ergo, subsequently continuously deformed to a circle) by switching the crossings for a finite number of times.
(To remind the readers, a crossing is defined as positive or negative according as the overpass must be rotated counter-clockwise or clockwise to bring it into coincidence with the underpass.)
One may note that the resulting circle may be converted back to the original knot simple by performing the operations in the reverse order.
With this in mind, let us consider a tubular region with the circle as axis. The cross-section of the tube is small and over that the velocity of field, which we suppose is filling the tube with strength $V$, is uniform; each streamline is, of course, a concentric circle to the circle serving as the axis.
Now, let us transversely cut the tube somewhere and reconnect it back after giving it a twist through an angle $2\pi N$ (where, $N$ is an integer).
This way we are introducing a stream-helicity of magnitude $NV^2$.
Then by introducing proper switching loops with similar strength, this construction may be changed to a knot with stream-helicity:
\be
S=[N+2(n_+-n_-)]V^2
\label{21}
\en
where, $n_+$ and $n_-$ are respectively the number of positive and negative switches needed to create the knot whose stream-helicity we are interested in.
One may prove that $N+2(n_+-n_-)$ is actually the linking number of any pair of streamlines in the knotted streamtube.
It also is the self-linking number of the `framed' knot which is `framed' using Frenet-Serret coordinate system\cite{5}.
Point to be noted is that for the kind of velocity field we are discussing $S$ will remain conserved and hence emerges as a topological invariant, for, evidently $S$ depends on the topology of the knotted streamtube.
\section{CONCLUSION}
To summarise, a new conserved quantity -- stream-helicity -- has been found (albeit, in a restricted sense) in fluid dynamics.
By seeking a topological interpretation for it in the certain configurations of linked and knotted streamtubes, the bridge between topology and fluid dynamics has been made even stronger.
In addition, as a byproduct, the seemingly non-physical quantity -- velocity vector potential -- has given itself a sort of physical meaning by getting involved in measuring the degree of knottedness of streamtubes.\\ \\
The author wants to thank Prof. J.K. Bhattacharjee for various important constructive comments on this work.
The author is indebted to Mr. Ayan Paul for providing him with various scholarly articles.
The anonymous referee is heartily thanked for suggesting the reference (\cite{Berger}).
CSIR (India) is gratefully acknowledged for the fellowship awarded to the author.


\end{document}